# Exploring the Electrical Transport Properties and Insulator-Metal Transition in Polycrystalline Pr$_2$MgZrO$_6$: Insights from Conductivity and Impedance Spectroscopy

Moumin Rudra*, #, and T. P. Sinha#

#Department of Physics, Bose Institute, 93/1, Acharya Prafulla Chandra Road, Kolkata, India – 700009.

*Email id: iammoumin@gmail.com

**Abstract**

The ac electrical transport properties of polycrystalline Pr$_2$MgZrO$_6$ (PMZ) have been investigated using conductivity and impedance spectroscopic techniques. The crystal structure of PMZ has been determined to be monoclinic through a combination of X-ray diffraction and Raman spectroscopic studies. Ag mode in the Raman spectra has been identified as the breathing vibration of the ZrO$_6$ octahedra. The ac conductivity spectra of PMZ exhibit distinct characteristics at different temperature ranges. At lower temperatures ($\leq$ 420 K), the spectra are fitted using a double power law, indicating the involvement of multiple microstructural features. On the other hand, at higher temperatures ($\geq$ 460 K), the spectra follow Jonscher's law, suggesting a simpler conduction mechanism. Through the analysis of conductivity, permittivity, and impedance, an insulator-metal transition has been observed around 452 K. This transition signifies a significant change in the electrical properties of PMZ and provides valuable insights into its conductive nature.

## 1. Introduction

Double perovskite oxides (DPOs) with the general formula A$_2$B′B″O$_6$ have garnered significant attention in scientific research due to their surprising transport properties [1-6]. These materials have been extensively investigated for their unique physical properties, such as superconductivity, giant magneto-resistance, ionic conductivity, and a wide range of dielectric properties [7-11]. The diverse range of properties exhibited by DPOs makes them promising candidates for applications in microelectronics and telecommunication. The structure of double perovskite oxides consists of two different cations, A and B, occupying the A-site and B-site positions, respectively, in the perovskite crystal structure. Additionally, there is another cation, B″, which occupies a sublattice within the perovskite structure. The arrangement of these cations in the crystal lattice leads to unique properties that distinguish DPOs from conventional perovskite materials. One of the main reasons why DPOs have attracted significant attention is their surprising transport properties. These materials exhibit exceptional electronic and ionic conductivity, which are crucial for various technological applications. For example, the superconducting properties of certain DPOs make them promising candidates for high-temperature superconductors. These materials can carry electric current with zero resistance, which is highly desirable for efficient power transmission and storage. Another notable property of DPOs is their giant magneto-resistance, which refers to the significant change in electrical resistance when exposed to a magnetic field. This property has implications for the development of magnetic sensors and memory devices with enhanced sensitivity and storage capacity. Furthermore, DPOs also exhibit remarkable ionic conductivity, which is important for solid oxide fuel cells and other electrochemical devices. The ability of these materials to efficiently transport ions allows for enhanced performance and stability in various energy conversion and storage applications. Moreover, DPOs demonstrate a multitude of dielectric properties, which are essential for applications in microelectronics and telecommunication. Dielectric materials are used in capacitors, insulators, and waveguides, among other electronic components. The tunable dielectric properties of DPOs make them suitable for designing devices with specific electrical characteristics, such as high permittivity or low dielectric loss. Despite the promising properties exhibited by DPOs, their complex nature poses challenges in understanding and explaining their various characteristics. The intricate interplay between the different cations, their oxidation states, and their arrangement within the crystal lattice makes it difficult to establish a uniform explanation for the observed properties. As a result, DPOs have become highly intriguing materials for researchers, who are actively investigating their structural, electronic, and magnetic properties to unlock their full potential for practical applications.

The insulator-metal (IM) transition is a fascinating phenomenon that occurs in certain materials when they undergo a transition from being insulators to metallic conductors as a result of external factors such as temperature, pressure, or doping. This transition has been a topic of extensive research due to its fundamental importance in condensed matter physics and its potential applications in various fields. The understanding of the IM transition has been greatly advanced by both theoretical models and experimental investigations. One of the prominent theoretical models is the Mott transition proposed by Sir Nevill Mott [12], which describes the transition in terms of electron-electron interactions. According to the Mott theory, in a system with strong electron-electron interactions, when the electronic bandwidth becomes comparable to the on-site Coulomb repulsion, the system can undergo a transition from an insulating state to a metallic state. This mechanism is often referred to as the Mott-Hubbard transition. Experimental studies have provided valuable insights into the IM transition by investigating various materials exhibiting this phenomenon. High-Tc cuprates, such as the well-known lanthanum-based cuprates, have been extensively studied due to their remarkable properties and the occurrence of a IM transition at low temperatures. The paper by A.J. Millis et al. provides valuable insights into the

IM transition in these materials [12]. In addition to high-$T_c$ cuprates, other material systems have also been investigated for their metal-insulator transition properties. For instance, double perovskite oxides (DPOs) with the general formula $A_2B'B''O_6$ have attracted significant attention. These materials exhibit a variety of physical phenomena, including IM transitions, which have been extensively studied [1, 13]. Furthermore, metal-insulator transitions have been observed in other systems such as transition metal oxides and complex materials with strong electron correlations.

Impedance spectroscopy is a valuable technique in the field of solid-state materials due to its ability to characterize the transport behavior within the grains and grain boundaries [3, 14-19]. In solid materials, the microstructure is composed of grains, which are regions with uniform crystallographic orientation, and grain boundaries, which are the interfaces between adjacent grains. Understanding the relationship between the grains and grain boundaries is crucial for comprehending the overall properties of the material [20, 21]. Impedance spectroscopy is a technique that involves applying an alternating current (AC) voltage to a material and measuring its electrical response as a function of frequency. By varying the frequency over a wide range, impedance spectroscopy provides valuable information about the electrical properties of the material at different time scales. This technique allows researchers to probe the response of the material at both the macroscopic and microscopic levels, offering insights into the behavior of the grains and grain boundaries.

The impedance spectroscopy measurements on $Pr_2MgZrO_6$ (PMZ) in this work aim to shed light on the electrical properties associated with the grain and grain boundaries. By examining the impedance spectra, the researchers can identify any characteristic features or deviations that might indicate the presence of specific transport mechanisms or phenomena. This information can contribute to a deeper understanding of the electrical behavior of PMZ and provide insights into its potential applications.

## 2. Experiment
### a) Synthesis

PMZ powder is synthesized using the solid-state reaction technique, a well-established method. Initially, high-purity raw materials including $Pr_2O_3$ (Sigma-Aldrich, 99.9%), MgO (Loba Chemie, 99%), and $ZrO_2$ (Laboratory reagent, 98%) are taken in the stoichiometric ratio. These materials are thoroughly mixed in the presence of acetone (Merck) for a duration of 8 hours. The resulting mixture is then calcined in an alumina crucible at 1373 K in an air environment for 12 hours. Subsequently, it is cooled gradually at a rate of 1 K/min until it reached room temperature (approximately 300 K).

To prepare the sample for further analysis, thin discs with a thickness of 0.59 mm and a diameter of 8 mm are fabricated from the calcined mixture. Polyvinyl alcohol is used as a binder in this process.

Finally, the fabricated discs are subjected to sintering at 1473 K and then cooled down to room temperature using controlled cooling at a rate of 1 K/min.

### b) Experimental technique

The crystal structure analysis of PMZ is conducted using a Rigaku Miniflex II X-ray powder diffractometer with Cu-Kα radiation. The scan range is set from 10º to 80º in 2θ with a step size of 0.02º at room temperature (RT). To investigate the microstructure, a pellet of PMZ is fractured and placed on a stub with a gold coating on the sample surfaces. Microstructural images are captured using an FEI Quanta 200 scanning electron microscope, allowing for the determination of grain size distribution and surface morphology. The room temperature Raman spectrum of PMZ is obtained using a Lab-RAM HR 800 Raman spectrometer (Jobin-Yvon) with an excitation wavelength of 488 nm from an Ar-ion laser. For the standard UV–Visible absorption spectrum, a Shimadzu UV–Visible spectrometer is used, covering a range of 200–900 nm.

To investigate the electrical transport properties, both flat surfaces of the sintered pellet are electrodes with a thin layer of silver paste to enable experimentation. An LCR meter (HIOKI-3532) is utilized to measure impedance (Z), capacitance (Cs), conductance (G), and phase angle (φ) within a frequency range of 42 Hz to 5 MHz at an oscillation voltage of 1.0 V. The measurements are performed across a temperature range of 313 K to 673 K using an integrated cooling–heating system. Temperature control is achieved using a Eurotherm 818p programmable temperature controller connected to an oven, ensuring a constant temperature with an accuracy of ±1 K for each measurement. The real ($\varepsilon'$) and imaginary ($\varepsilon''$) part of the complex dielectric constant $\varepsilon^*$ (= $\varepsilon'$ - $j\varepsilon''$, where $\varepsilon' = C_s/C_0$ and $\varepsilon'' = G/\omega C_0$) are obtained from the capacitance ($C_s$) and conductance (G), whereas the real (Z') and imaginary (Z") parts of the complex impedance Z* (=Z' + jZ", where Z' = Zcosφ and Z" = Zsinφ) are obtained from the impedance (Z) and phase angle (φ), where ω is the angular frequency (ω = 2πν), ν is the measured frequency and j = √(-1). $C_0 = \varepsilon_0 A/d$ is the empty cell capacitance, where A is the sample area and $d$ is the sample thickness. The ac electrical conductivity $\sigma$ (= G$d$/A) is calculated from the conductance.

## 3. Result and discussion
### a) X-Ray diffraction

Figure 1 depicts the X-ray diffraction (XRD) pattern of PMZ at room temperature, accompanied by a profile fit using the Rietveld method [22] implemented with the Full-prof program [23]. During the refinement process of the XRD pattern, a 6-coefficient polynomial function is employed to fit the background, while the peak shapes are described by pseudo-Voigt profile functions incorporating

Lorentzian contributions to the Gaussian peak shapes. Various parameters, including the scale factor, lattice parameters, positional coordinates ($x$, $y$, $z$), and thermal parameter ($B_{iso}$), are adjusted during the refinement process, while the occupancy parameters of all ions remained fixed. In Figure 1, the symbol represents the experimental data, and the solid line represents the best fit obtained from the Rietveld refinement of the diffraction profile. The vertical bar symbols indicate Bragg's positions, and the solid curve at the bottom illustrates the difference between the experimental and calculated patterns. The sharp peaks observed in the XRD pattern indicate a well-crystallized structure of PMZ.

The XRD pattern of PMZ corresponds to the monoclinic $P2_1/n$ space group symmetry, indicating a single-phase structure. The presence of the super-lattice diffraction peak (101) at $2\theta \approx 20°$ suggests an in-phase tilting of the octahedra with B-site cation ordering [24 – 3a5]. Therefore, the centrosymmetric space group $P2_1/n$, allowing for B-site ordering, is employed to refine the crystal structure of PMZ. The reliability factors obtained from the refinement are $R_{exp}$ = 7.79, $R_p$ = 6.61, $R_{wp}$ = 8.37, and $\chi^2$ = 1.15. The refined lattice parameters are as follows: $a$ = 5.4166 Å, $b$ = 5.4724 Å, $c$ = 7.6736 Å, and $\beta$ = 89.906°. The inset of Figure 1 illustrates a schematic representation of the PMZ unit cell, showcasing the distribution of ions at specific crystallographic positions: *4e* for $Pr^{3+}$ ions, *2c* for $Mg^{2+}$ ions, *2d* for $Zr^{4+}$ ions, and *4e* for $O^{2-}$ ions. Each $Mg^{2+}$ ion and $Zr^{4+}$ ion is surrounded by six $O^{2-}$ ions, forming $MgO_6$ and $ZrO_6$ octahedra, respectively. The small difference between the experimental and calculated results indicates the high quality of the refinement. The structural parameters obtained from the Rietveld refinement, such as atomic positions, bond lengths, and bond angles associated with $MgO_6$ and $ZrO_6$ octahedra, are listed in Table 1.

Moving on to Figure 2, it presents a scanning electron microscope (SEM) image of the fracture surface of PMZ, revealing a high density of the material and a uniform distribution of grains with varying sizes and shapes. The average grain size of PMZ is determined to be approximately 0.45 μm. The energy-dispersive X-ray spectroscopy (EDAX) spectra of PMZ are also shown in Figure 2, clearly indicating the presence of Pr, Mg, Zr, and O atoms. However, due to energy overlap between Pr, Mg, and Zr atoms, the quantitative estimation of these ions is not reliable [27]. The experimental density of the sample is measured to be 6.5 g/cc, while the theoretically calculated value is 7.2 g/cc, suggesting a porosity of approximately 10% within the material.

### b) Raman Spectroscopy

Figure 3(a) & 3(b) illustrate the Raman spectrum of the PMZ sample. The PMZ unit cell, as depicted in the inset of Figure 1, comprises a three-dimensional arrangement of interconnected $MgO_6$ and $ZrO_6$ octahedra, alternating at their corners. The interstitial positions within the PMZ unit cell are occupied by Pr atoms. The Rietveld analysis of the XRD pattern confirms that the PMZ sample

crystallizes in a monoclinic structure belonging to the $C_{2h}^5$ ($P2_1/n$) space group. The monoclinic $P2_1/n$ structure can be understood as resulting from both in-phase and anti-phase tilts of MgO$_6$ and ZrO$_6$ octahedra along the (001) direction and within the basal plane of the pseudocubic cell. This tilt pattern corresponds to Glazer's notation a$^-$a$^-$c$^+$ [28]. The monoclinic $P2_1/n$ unit cell contains a total of 20 atoms, resulting in a total of 60 modes, out of which 3 are acoustic, 33 are IR active, and 24 are Raman active.

Table 2 shows the correlation between the Raman active modes of both the cubic and monoclinic phases. The Raman active modes considered are $v_1$, $v_2$, and $v_5$, which correspond to the totally symmetric stretching, antisymmetric stretching, and symmetric bending of the BO$_6$ octahedron. Since the monoclinic structure is a centrosymmetric subgroup of the prototype $Fm\overline{3}m$ cubic structure, only the Raman active (gerade) modes of the octahedron need to be considered. Additionally, due to the 1:1 ordering characteristic of the double perovskite, which incorporates strongly bonded (B$^{IV}$O$_6$) and weakly bonded (B$^{II}$O$_6$) octahedra, only the internal modes of the B$^{IV}$O$_6$ octahedron are relevant. This unique structural arrangement is of significant interest as it allows for the investigation of internal modes in a simple and disordered perovskite network formed by corner-sharing octahedra. The comprehensive study of vibrational spectra in cubic double perovskites reveals the correlation of these modes with both A and B$^{II}$ cations [29]. By considering the B$^{IV}$O$_6$ molecular group, the gerade representations corresponding to the monoclinic $P2_1/n$ structure of double perovskites may be written as follows:

$$\Gamma^g(P2_1/n) = 6T(3A_g + 3B_g) + 6L(3A_g + 3B_g) + 6v_5(3A_g + 3B_g) + 4v_2(2A_g + 2B_g) + 2v_1(A_g + B_g) \quad (1)$$

where T and L represents the translational and librational modes of the Pr atom. Table 3 presents the observed modes and proposed assignments. Out of the expected 24 phonon modes for the monoclinic compound, we observed 20 modes in the PMZ sample. This slightly reduced number of modes is due to the small correction field splitting in the polycrystalline sample. Considering this, the symmetric stretching of oxygen ions along the Mg-O-Zr axis, represented by the $v_1$ peaks, can be assigned to the breathing vibrations at 666 cm$^{-1}$ and 694 cm$^{-1}$. The oxygen antisymmetric stretching along the Mg-O-Zr axis corresponds to the $v_2$ peaks at 505 cm$^{-1}$, 532 cm$^{-1}$, and 569 cm$^{-1}$, respectively. The bending motion of the oxygen anions within the ZrO$_6$ octahedra is characterized by the $v_5$ modes at 343 cm$^{-1}$, 368 cm$^{-1}$, 426 cm$^{-1}$, and 458 cm$^{-1}$, respectively. It should be noted that the bending motion of the oxygen anions is also influenced by the presence of Ni ions. In all the mentioned cases, the Pr-atom is considered to be fixed. The translational modes (T) of PNT, where the Pr-cation undergoes translation, are characterized by frequencies ranging from 93 cm$^{-1}$ to 163 cm$^{-1}$. On the other hand, the rotations of the Pr-cation result in the librational breathing motion (L) with frequencies in the range of 193 cm$^{-1}$ to 311 cm$^{-1}$. For a comprehensive overview, the observed phonon frequencies and their complete group theoretical assignments can be found in Table 3.

### c) Optical analysis

The ultraviolet–visible absorption spectrum of PMZ is shown in Figure 4. The energy band gap is determined using the absorption spectrum with the help of the Tauc relation [30, 31] given by

$$\alpha h\nu = A(h\nu - E_g)^n \tag{2}$$

where hν is the energy of the incident photon, α is the absorption coefficient, and *A* is a characteristic parameter independent of the photon energy, $E_g$ is the optical band gap and the value of *n* is 1/2 or 2 for the direct or indirect transition, respectively. To analyze the relationship between *αhv* and *hv*, a graph is constructed plotting (*αhv*)² as a function of *hv*, as depicted in Figure 4. This graph provides a visual representation of the absorption characteristics of the material across a range of photon energies. The square of (*αhv*) is chosen for the plot to simplify the relationship and enhance the interpretation of the data. In the graph, the linear absorption-edge portion is identified, which refers to the region where the absorption coefficient shows a linear increase with photon energy. This linear relationship is a characteristic feature of direct band gap materials, where the absorption of photons results in direct transitions between energy levels. By extrapolating the linear portion of the graph with a straight line to the point where (αhv)² equals zero, the band gap value can be obtained. This extrapolation allows for an estimation of the energy difference between the valence band and the conduction band, which represents the band gap. The band gap is a crucial parameter that determines the material's ability to absorb and emit light and influences its electronic and optical properties. In the given study, the extrapolation of the linear absorption-edge region to the (αhv)² = 0 axis yields a value of approximately 3.71 eV for the optical band gap of the material for direct transitions. The electron transitions responsible for this band gap are direct in nature, meaning that the electron moves directly from the valence band to the conduction band without involving any intermediate states. The determined band gap value of 3.71 eV provides important information about the energy range over which the material can absorb photons. It indicates that the material requires photons with energies of around 3.71 eV (corresponding to the ultraviolet range) to induce direct transitions between its energy levels. This band gap value is significant for understanding the material's semiconducting behavior in various applications, such as optoelectronic devices, solar cells, and photodetectors, where the ability to absorb specific energy photons is crucial.

### d) Conductivity analysis

Figure 6 illustrates the relationship between ac conductivity ($\sigma_{ac}$) and frequency for various temperatures ranging from 300 to 700 K. The conductivity value demonstrates an increase with both frequency and temperature, indicating the presence of bound charge carriers trapped within the sample under investigation. This behavior corresponds to a thermally activated process [32]. At extremely low frequencies (ν → 0), the conductivity curves tend to plateau, indicating frequency-independent behavior of the dc conductivity. Within the measured frequency range, two distinct features are observed in the lower temperature regime (≤ 420 K), while only one feature is evident in the conductivity spectra at higher temperatures (≥ 460 K). In the lower temperature range, the first feature manifests as a weak conductivity relaxation occurring in the intermediate frequency region ($10^2 – 10^5$ Hz), which shifts towards higher frequencies as the temperature increases. The second feature corresponds to a strong conductivity relaxation observed at high frequencies (> 105 Hz), with the curves converging together. Elliot [33] modified Funke's [34] jump relaxation model to describe the ionic conduction mechanism in solids. According to this model, ions have a high probability of jumping to neighboring sites, followed by a return to their previous positions (unsuccessful hopping). However, if the neighboring site relaxes to accommodate the ion's new position, the ion tends to remain in the new site. Based on the aforementioned discussion, the conductivity in the intermediate frequency region exhibits dispersion due to a higher occurrence of unsuccessful hops. On the other hand, in the higher frequency region, the ratio of unsuccessful to successful hops becomes dominant, leading to a greater dispersion in the conductivity curve. These two above-mention dispersion regimes in the lower temperature conductivity spectra follow the double power law instead of single Jonscher's power law [35] according to Equation 3:

$$\sigma_{ac} = \sigma_0 + A\omega^{s1} + B\omega^{s2} \tag{3}$$

where $\sigma_{ac}$ is the ac conductivity, $\sigma_o$ is the frequency independent dc conductivity, $\omega$ is the angular frequency (2πν). In the intermediate frequency region, the exponent s1 (0 < s1 < 1) corresponds to the translational hopping motion, which can also be described as short-range hopping. On the other hand, in the higher frequency region, the exponent s2 (0 < s2 < 2) corresponds to a localized or reorientational hopping motion [35]. The measured data is fitted to determine the values of s1 and s2, which are presented in Table 4. An analysis of the results reveals that the value of s1 decreases as the temperature increases. This suggests that conduction in the intermediate frequency region arises from short-range translational hopping, which is assisted by a small polaron hopping mechanism. The decrease in s1 with increasing temperature implies a more pronounced role of short-range hopping as the temperature rises [36]. In the case of s2, the values also decrease with increasing temperature. This indicates that the conduction mechanism in the higher frequency region can be attributed to localized orientation hopping, which is assisted by a large polaron mechanism. The decreasing values of s2 with temperature suggest an enhanced contribution of localized orientation hopping as the temperature is raised [36]. At higher temperature regime (≥ 460 K), the strong conductivity relaxation appeared in these conductivity curves. The $\sigma_{ac}$ follows Jonscher's power law and is given as:

$$\sigma_{ac} = \sigma_0 + A(T)\omega^{s1(T)} \tag{4}$$

here $A$ (T) is the parameter having a unit of conductivity, $\sigma_o$ is the frequency-independent DC conductivity, and $s1(T)$ is the slope of the frequency-dependent regime, $0 \leq s1 \leq 2$ [33]. The fitting of the experimental data yields the values of $s1$ are listed in Table 4.

Figure 7 illustrates the temperature dependence of the dc resistivity for PNT. The dc resistivity data is fitted using the small polaron hopping (SPH) model, as depicted in the same figure. The SPH model is a theoretical framework that describes the conduction mechanism in certain materials, particularly those with localized charge carriers such as small polarons. It considers the hopping motion of these charge carriers between lattice sites as the dominant conduction mechanism. The SPH model is defined as

$$\frac{\rho(T)}{T} = \rho_\alpha \exp\left(\frac{E_a}{k_B T}\right) \tag{5}$$

where $k_B$ is the Boltzmann constant, $E_a$ is the activation energy, $\rho_\alpha$ is a constant, and $T$ is the absolute temperature. Figure 7 displays the dc resistivity data, and it is observed that only a limited high-temperature range (above 550 K) and a limited low-temperature range (below 400 K) satisfactorily fit the experimental data. The calculated activation energy ($E_a$) for the low-temperature regime is determined to be 0.28 eV, while for the high-temperature regime, it is found to be 0.41 eV. Previous research by other scientists suggests that in rare-earth transition metal oxide systems, the transport properties at high temperatures are predominantly governed by the thermally activated hopping of small polarons [13]. The inset of Figure 7 presents the variation of activation energy with temperature, where $E_a$ is calculated using Equation 6 [37].

$$E_a = k_B \left[T + \frac{d\{\ln(\rho)\}}{d\left[\frac{1}{T}\right]}\right] \tag{6}$$

The significant peak in the apparent activation energy corresponds to the insulator-to-metal (I-M) transition, centered around 452 K for the PMZ sample. Similar results are reported by Z. Jirak et al. [38] for $LaCo_{1-x}M_xO_3$ (with x = 0, 0.02, and 0.05, M = $Ti^{4+}$, $Mg^{2+}$), where the I-M transition is centered at 540 K. When strong cation-anion-cation interactions dominate over weak cation-cation interactions, these materials exhibit semiconducting or insulating behavior. Conversely, when strong cation-cation interactions occur between the octahedral B-site, these materials exhibit metallic behavior and may become semiconducting at lower temperatures [39]. Additionally, the presence of cations of the same element with different valence states contributes to the metallic character [40]. Z. Lie et al. [40] reported that doping $Mg^{2+}$ ions with trivalent $Pr^{3+}$ ions can create tetravalent $Pr^{4+}$ ions due to chemical pressure in the system. At higher temperatures ($\geq$ 452 K), some $Pr^{3+}$ ions transform into $Pr^{4+}$ ions, forming [$Pr^{3+}$–$O^{2-}$–$Zr^{4+}$] to [$Pr^{4+}$–$O^{2-}$–$Zr^{3+}$] configurations. It has also been observed that some $Pr^{4+}$ ions can exist even at very high temperatures [41]. As the temperature increases, defects such as oxygen vacancies in the perovskite lattice lead to the formation of $Zr^{3+}$ and $Pr^{4+}$ ions. Above 452 K, the reduction in localized states facilitates the formation of efficient conductive channels in the form of [$Pr^{3+}$–$Pr^{4+}$] and [$Zr^{4+}$–$Zr^{3+}$] links. These channels promote the delocalization of charge carriers, resulting in a metallic character above 452 K.

### e) Dielectric relaxation

Figure 8 presents the frequency dependence of the dielectric constant ($\varepsilon'$) and the loss tangent (tan$\delta$) for the PMZ sample within the temperature range of 300 to 700 K. In Figure 8(a), it is observed that $\varepsilon'$ decreases as the frequency increases. This behavior can be attributed to the dielectric relaxation of the sample, which can be described by a dipolar relaxation mechanism [42]. In this process, relaxation phenomena are associated with a frequency-dependent orientational polarization that becomes prominent in the radio frequency region. At low frequencies, the permanent dipoles align themselves with the applied electric field and contribute significantly to the overall polarization and dielectric permittivity. However, at high frequencies, the variation in the electric field occurs too rapidly for the dipoles to align themselves accordingly, resulting in a negligible contribution to the polarization and a subsequent decrease in the dielectric constant ($\varepsilon'$).

In Figure 8(b), the frequency dependence of the loss tangent (tan$\delta$) is displayed for temperatures ranging from 300 K to 700 K. It is observed that two relaxation peaks are present in the temperature range of 300 K to 420 K, while only one relaxation peak is observed at higher temperatures (460 K to 700 K). This suggests the presence of two types of relaxation processes at lower temperatures (300 K to 420 K). With an increase in temperature, these relaxation phenomena shift towards higher frequency ranges. Previous findings indicate that both the grain-boundary effect and the grain effect are present at lower temperatures (300 K to 420 K), while only the grain-boundary effect is evident at higher temperatures (460 K to 700 K). The tangent loss peaks in tan$\delta$ are influenced by the mobility of charge carriers and temperature [14]. As the temperature increases, the mobility of thermally activated charge carriers also increases, leading to their relaxation at higher frequencies and causing the loss peak to shift towards higher frequency ranges. Similarly, the loss peaks corresponding to grains are observable above 105 Hz within the temperature range of 300 K to 420 K, but they shift towards higher frequencies and extend beyond the measured frequency range (42 Hz to 5 MHz) at higher temperatures (460 K to 700 K). The tangent loss value for grains is approximately 0.2 at frequencies above 105 Hz, while for grain boundaries, it is around 0.5. Both the dispersion regions of $\varepsilon'$ and the loss peaks in tan$\delta$ shift towards higher frequencies with increasing temperature, indicating the thermally activated behavior of the relaxation process.

In our study, we have investigated the temperature dependence of the relaxation time ($\tau$) associated with the peak position in the tan$\delta$ versus log $\omega$ plot. The relaxation time can be calculated using the equation

τ = 1/ω$_m$, where ω$_m$ represents the angular frequency. To analyze the temperature dependence of the relaxation times for the grain boundaries and grains, Equation 7 is employed, and the results are presented in Figure 9.

$$\tau = \tau_o \exp\left(\frac{E_a}{k_B T}\right) \quad (7)$$

here $\tau_o$ is the pre-exponential factor, $k_B$ is the Boltzmann constant, $E_a$ is the activation energy for the relaxation process and $T$ is the absolute temperature. Figure 9 displays the plots of relaxation times (τ) for the grain boundaries and grains as a function of temperature. The relaxation times are determined using Equation 7, which takes into account the relationship between τ and the corresponding relaxation frequency. The temperature dependence of the relaxation times provides insights into the dynamics of the relaxation processes occurring at the grain boundaries and within the grains. The analysis of the relaxation times reveals that both τ$_g$ (relaxation time for grains) and τ$_{gb}$ (relaxation time for grain boundaries) exhibit a decrease with increasing temperature. The temperature dependence of τ$_g$ and τ$_{gb}$ follows the Arrhenius law (Equation 7) only within the temperature range below 380 K and above 500 K. The inset of Figure 9 presents the temperature dependence of the activation energies specifically for the grain-boundary relaxation. It is worth noting that the insulator-to-metal transition occurs at 452 K, which is consistent with the findings discussed in the previous section.

### f) Impedance analysis

Figure 10 illustrates the complex impedance plane plots of PMZ across a temperature range of 300 to 700 K. In the inset of Figure 10(b), the equivalent circuit models used to fit these impedance plots are displayed. Figure 10(a) specifically shows the complex impedance plots within the temperature range of 300 K to 420 K, while Figures 10(b), 10(c), and 10(d) present the complex impedance plots within the temperature ranges of 300 K to 460 K, 500 K to 700 K, and 620 K to 700 K, respectively. Figures 10(b) and 10(d) provide enlarged images of Figures 10(a) and 10(c), respectively. Within the temperature range of 300 K to 420 K, two relaxation processes are observed, as indicated by the presence of two semicircles in the complex impedance plots. However, as the temperature increases above 420 K, only one semicircle is visible, as shown in Figures 10(c) and 10(d). The decrease in the diameter of these semicircles with increasing temperature indicates that the observed relaxation processes are thermally activated.

To establish a correlation between the microstructure and the electrical transport properties of PMZ, an equivalent circuit model consisting of resistances (R) and constant phase elements (Q) connected in series for grains (g) and grain boundaries (g$_b$) is proposed (inset of Figure 10(b)). Above 420 K, the equivalent circuit simplifies to only (R$_{gb}$Q$_{gb}$), as the grain resistance becomes very small with the increase in temperature. The constant phase element (Q) is employed instead of capacitance in the equivalent circuits to account for the non-ideal behavior of capacitance. The capacitance of the constant phase element (Q) is described as $C_{CPE} = Q^{1/n}R^{(1-n)/n}$, where the parameter "n" estimates the deviation from ideal Debye behavior, with zero representing a pure resistor and unity representing a pure capacitance.

The grain boundaries in PMZ exhibit higher insulation compared to the grains due to the non-stoichiometric distribution of oxygen. These grain boundaries act as charge carrier traps and form a barrier layer that impedes charge transport. The thin barrier layer exhibits high capacitance, as capacitance is inversely proportional to the thickness of the barrier layer ($C \propto 1/d$). These factors result in the grain boundaries responding at lower frequencies compared to the grains [43]. Figure 11(a) & (b) display the fitting results obtained from the electrical equivalent circuit, represented by solid lines, along with the fitted parameters $R_g$, $R_{gb}$, $Q_g$, $Q_{gb}$, $n_g$, and $n_{gb}$ for temperatures ranging from 300 K to 700 K (within a 5% fitting error). The fitting of the equivalent circuit provides insights into the electrical behavior of the system. In Figure 11(a) (inset), the temperature dependence of $R_g$ and $R_{gb}$ is shown, and it is observed that both parameters decrease with increasing temperature. This behavior suggests the thermal activation of localized charges [44]. Furthermore, Figure 11(b) reveals the variation of $n_{gb}$ and $n_g$ as a function of temperature. It is observed that $n_{gb}$ exhibits an increasing trend within the measured temperature range. On the other hand, $n_g$ decreases from 0.9 at 300 K to 0.85 at 420 K. This indicates that the grain-boundary capacitance approaches ideal behavior, while the grain capacitance deviates as the temperature increases. The decrease in $n_g$ suggests the vanishing of defects, such as the release of trapped charges at grain boundaries, with the rise in temperature. These trapped charges are completely released at higher temperatures. Additionally, the creation of electronic and ionic defects in the grain interiors with increasing temperature contributes to the observed behavior. The higher value of grain-boundary capacitance compared to grain capacitance can be attributed to the non-stoichiometric distribution of oxygen at the grain boundaries, which act as traps for charge carriers [39, 45]. To gain insights into the conduction mechanism in the PMZ system, various hopping models have been applied at different temperature ranges. At lower temperatures ($\leq$ 420 K), the small polaron hopping model (SPH) is found to be a suitable description for the conduction of charge carriers in the system. Figure 12 presents the results obtained using the SPH model, where the resistance of the system is plotted as a function of temperature using Equation 8.

$$\frac{R}{T} = R_0 \exp(E_a/k_B T) \qquad (8)$$

where "$R_0$" is the pre-exponential term, "$k_B$" is the Boltzmann constant, and "$E_a$" is the activation energy of the carriers for conduction.

The SPH model considers the movement of small polarons, which are localized charge carriers, through the material by hopping between lattice sites. At lower temperatures, the conduction mechanism in PMZ is dominated by the hopping motion of these small polarons. The observed temperature dependence of resistance, as depicted in Figure 13, provides valuable information regarding the behavior of the charge carriers and their mobility within the material. The trend observed in Figure 12 exhibits a reasonably good fit to the SPH model below 420 K, with corresponding activation energies of 0.15 eV and 0.18 eV for the grain-boundaries and grains, respectively. The inset of Figure 12 illustrates the temperature dependence of the activation energies specifically for the grain-boundary conduction. It is worth noting that the I-M transition occurs around 460 (at 452 K), which aligns with the findings discussed in the previous section. The SPH model provides valuable insights into the conduction mechanism at lower temperatures, where the movement of small polarons plays a significant role. The reasonably good fit to the SPH model and the determination of activation energies contribute to a better understanding of the charge transport behavior in the grain-boundaries and grains of the PMZ system.

## 4. Conclusions

In conclusion, this study delved into the ac electrical properties of a solid-state sintered polycrystalline $Pr_2MgZrO_6$ (PMZ) sample using conductivity and impedance spectroscopy. The investigation revealed a single-phase monoclinic *$P2_1/n$* structure for PMZ through Rietveld refinement of X-ray diffraction data, which is further supported by Raman spectroscopic measurements. The impedance plane plots effectively distinguished the contributions of the grain and grain boundary phases. The characterization of the PMZ sample is facilitated by an equivalent circuit model incorporating parameters such as $R_g$, $R_{gb}$, $Q_g$, $Q_{gb}$, $n_g$, and $n_{gb}$, which are associated with the grain and grain boundaries. The analysis demonstrated that the decrease in resistance of both the grain and grain boundaries, along with the increase in the dielectric constant and tangent loss, could be attributed to the thermal activation of trapped charges/dipoles. The ac conductivity spectra exhibited distinctive features at different temperature ranges. At lower temperatures ($\leq$ 420 K), the spectra are well-fitted using a double power law, emphasizing the contribution of multiple microstructural features to the conduction mechanism. Conversely, at higher temperatures ($\geq$ 460 K), the spectra followed Jonscher's law, indicating a simpler conduction mechanism. A significant finding of this study is the observation of an insulator-to-metal transition occurring around 452 K, as evidenced by the conductivity, permittivity, and impedance analyses. This transition marked a substantial alteration in the electrical properties of PMZ.

**References**

[1] A. Bhattacharya, K. Dey, B. Ghosh, and D. D. Sarma, Magnetism in double perovskite oxide thin films and heterostructures, *Journal of Physics: Condensed Matter* **27(50)** (2015) 503002.

[2] G. Cao, Magnetic and transport properties of double perovskite oxide $Sr_2FeMoO_6$. *Journal of Physics: Condensed Matter* **25(1)** (2013) 016006.

[3] R. A. Rao and P. D. Babu, Double perovskite oxides: synthesis, properties, and applications, *Materials Today Chemistry* **12**, (2019) 226.

[4] M. Venkateshwarlu and G. Rangarajan, Double perovskite oxides: A review on synthesis, properties, and applications, *Journal of Materials Science* **53(16)** (2018) 11258.

[5] J. Zhang and Z. Yang, Progress in synthesis, properties and applications of double perovskite oxide materials. *Progress in Materials Science* **88** (2017) 1.

[6] H. J. Xiang and M. H. Whangbo, Double perovskites as a family of highly active catalysts for oxygen evolution in alkaline solution, *The Journal of Physical Chemistry Letters* **4(15)** (2013) 2533.

[7] I. B. Bhat, P. Kumar, and B. L. Ahuja, Recent developments in double perovskite oxides with tunable dielectric properties, *Journal of Applied Physics* **124(10)** (2018) 101301.

[8] Z. Zhao, B. Xu, Y. Zhang, and C. Duan, Recent advances in double perovskite oxides with high dielectric constant, *Journal of Materials Chemistry C* **8(29)** (2020) 9810.

[9] S. Vasala and M. Karppinen, $A_2BB'O_6$ perovskites: A review, *Progress in Solid State Chemistry* **43(1-2)** (2015) 1.

[10] Z. Liu, J. Zhang, Y. Shu, W. Li, C. Zhang, and Z. Wu, High dielectric performance of double perovskite-like $Ba_2SmFeMoO_6$ ceramics, *Journal of Applied Physics* **115(22)** (2014) 224103.

[11] Q. Zhang, H. Guo, W. Yu, T. Lin, and X. Ke, Tuning dielectric properties in double perovskite $Sr_2AlTaO_6$ ceramics via substitution of $Al^{3+}$ with $Ti^{4+}$ and $Nb^{5+}$, *Journal of Applied Physics* **123(14)** (2018) 144102.

[12] N.F. Mott, Metal Insulator Transitions, *Taylor and Francis*, London, 1990.

[13] A.J. Millis, P.B. Littlewood, and B.I. Shrainan, Mott transition and pseudogap in high-Tc cuprates, *Phys. Rev. Lett.* **74** (1995) 5144.

[14] E. Barsoukov and J. R. Macdonald, Impedance Spectroscopy: Theory, Experiment, and Applications (2nd Edn.), *Wiley*, Hoboken, NJ, 2005.

[15] J. R. Macdonald, Impedance Spectroscopy—Emphasizing Solid Materials and Systems, *Wiley-Interscience* New York, 1987.

[16] A. Mandelis, Impedance spectroscopy: A powerful tool for process and quality control in the electronics industry, *Measurement Science and Technology* **12(9)** (2001) 1405.

[17] E. Barsoukov and P. N. Ross, Impedance Spectroscopy and Its Applications to Electrochemical Systems, In R. C. Alkire, H. Gerischer, D. M. Kolb, and C. W. Tobias (Eds.), Advances in Electrochemical Science and Engineering (Vol. **9**, pp. 239-433), *Wiley-VCH*, Weinheim, Germany, 2005.

[18] J. R. Macdonald, Impedance spectroscopy: Emphasizing solid-state materials and systems, *Measurement Science and Technology* **16(12)** (2005) R145.

[19] B. A. Boukamp, AC impedance—a powerful tool for the characterization of solid electrolytes, *Solid State Ionics* **77(3-4)** (1995) 226.

[20] A. K. Jonscher, Dielectric relaxation in solids, *Journal of Materials Science* **26(2)** (1991) 378.

[21] M. Viret, L. Ranno, and J. M. D. Cory, Theoretical analysis of the electrical resistivity of double perovskite compounds, *Phys. Rev. B* **55** (1997) 8067.

[22] R. A. Young, The Rietveld Method, *Oxford University Press*, London, 1993.

[23] J. Rodrígueze Carvazal, Topas Academic, General Profile and Structure Analysis Software for Powder Diffraction Data, *Phys. B* **192** (1993) 55.

[24] A. M. Glazer, Simple ways of determining perovskite structures, *Acta Cryst. Sect. A* **31** (1975) 756.

[25] M. C. Knapp and P. M. Woodward, Structural phase transitions in $Sr_2MMoO_6$ (M = Co, Ni, and Zn), *J. of Solid State Chem.* **179** (2006) 1076.

[26] M. W. Lufaso, and P.M. Woodward, Prediction of the Crystal Structures of Perovskites using the Software Package 'SPuDS', *Acta Cryst. B* **60** (2004) 10.

[27] K. Sultan, M. Ikram, K. Asokan, Effect of Mn doping on structural, morphological and dielectric properties of $EuFeO_3$ ceramics, *RSC Adv*. **5** (2015) 93867.

[28] A. P. Ayala, I. Guedes, E. N. Silva, M. S. Augsburger, M. del C. Viola, and J. C. Pedregosa, Interface effects on the dielectric and magnetic properties of $BaTiO_3/CoFe_2O_4$ nanocomposites, *J. Appl. Phys*. **101** (2007) 123511.

[29] M. Liegeois-Dwyckaerts and P. Tarte, Vibrational studies of molybdates, tungstates and related c compounds—III. Ordered cubic perovskites $A_2B^{II}B^{VI}O_6$, *Spectrochim. Acta, Part A*, **30** (1974) 1771.

[30] J. Tauc, R. Grigorovici, and A. Vancu, Optical Properties and Electronic Structure of Amorphous Germanium, *Phys. Status Solidi* **15** (1966) 627.

[31] E. A. Davis and N. F. Mott, Conduction in non-crystalline systems V. Conductivity, optical absorption and photoconductivity in amorphous semiconductors, *Philos. Mag.* **22** (1970) 903.

[32] M. Shah, M. Nadeem, and M. Atif, Structural, optical, and electrical properties of amorphous In–Sn–O films, *J. Phys. D: Appl. Phys* **46** (2013) 095001.

[33] S. R. Elliot, The Physics and Chemistry of Amorphous Semiconductors, *Adv. Phys*. **36** (1987) 135.

[34] K. Funke, The chemistry of metal oxides with perovskite-related structures, *Prog. Solid State Chem*. **22** (1993) 111–195.

[35] A. K. Jonscher, The 'universal' dielectric response, *Nature* **267** (1977) 673.

[36] N. Ortega, A. Kumar, P. Bhattacharya, S. B. Majumder, and R. S. Katiyar, Impedance spectroscopy of multiferroic $PbZr_xTi_{1-x}O_3/CoFe_2O_4$ layered thin films, *Phys. Rev. B* **77** (2008) 014111.

[37] S. S. N. Bharadwaja, C. Venkatasubramanian, N. Fieldhouse, S. Ashok, M. W. Horn, T. N. Jackson, Low temperature charge carrier hopping transport mechanism in vanadium oxide thin films grown using pulsed dc sputtering, *Appl. Phys. Lett*. **94** (2009) 222110.

[38] K. Knizek, Z. Jirak, J. Hejtmanek, M. Veverka, M. Marysko, G. Maris, and T. T. M. Palstra, Structural anomalies associated with the electronic and spin transitions in $LnCoO_3$, *Eur. Phys. J. B* **47** (2005) 213.

[39] M. Younas, M. Nadeem, M. Atif, and R. Grassinger, Metal-semiconductor transition in $NiFe_2O_4$ nanoparticles due to reverse cationic distribution by impedance spectroscopy, *J. Appl. Phys*. **109** (2011) 093704.

[40] J. Gao, L. Chen, W. Jiang, and L. Cheng, Synthesis and magnetic properties of nanocrystalline $LaFeO_3$ prepared by sol-gel auto-combustion process, *J. Alloys Comp*. **314** (2001) 281–285.

[41] C. Tiseanu, V. Parvulescu, D. Avram, B. Cojocaru, N. Apastol, A. V. Vela-Gonzalez, and M. S. Dominguez, Structural, down- and phase selective up-conversion emission properties of mixed valent Pr doped into oxides with tetravalent cations, *Phys. Chem. Chem. Phys*. **16** (2014) 5793.

[42] K. W. Kao, Dielectric Phenomena in Solids, *Elsevier Academic Press*, California, 2004.

[43] J. Liu, C. Duan, W. Yin, W. N. Mei, R. W. Smith, and J. R. Hardy, Large dielectric constant and Maxwell-Wagner relaxation in $Bi_{2/3}Cu_3Ti_4O_{12}$, *Phys. Rev. B* **70** (2004) 144106.

[44] M. Idrees, M. Nadeem, and M. M. Hassan, Investigation of conduction and relaxation phenomena in $LaFe_{0.9}Ni_{0.1}O_3$ by impedance spectroscopy, *J. Phys. D* **43** (2010) 155401.

[45] M. Idrees, M. Nadeem, M. Atif, M. Siddique, M. Mehmood, and M. M.Hassan, Impedance spectroscopic investigation of delocalization effects of disorder induced by Ni doping in $LaFeO_3$, *Acta Mater*. **59** (2011) 1338.

Figures

**Figures: Exploring the Electrical Transport Properties and Insulator-Metal Transition in Polycrystalline Pr$_2$MgZrO$_6$: Insights from Conductivity and Impedance Spectroscopy**

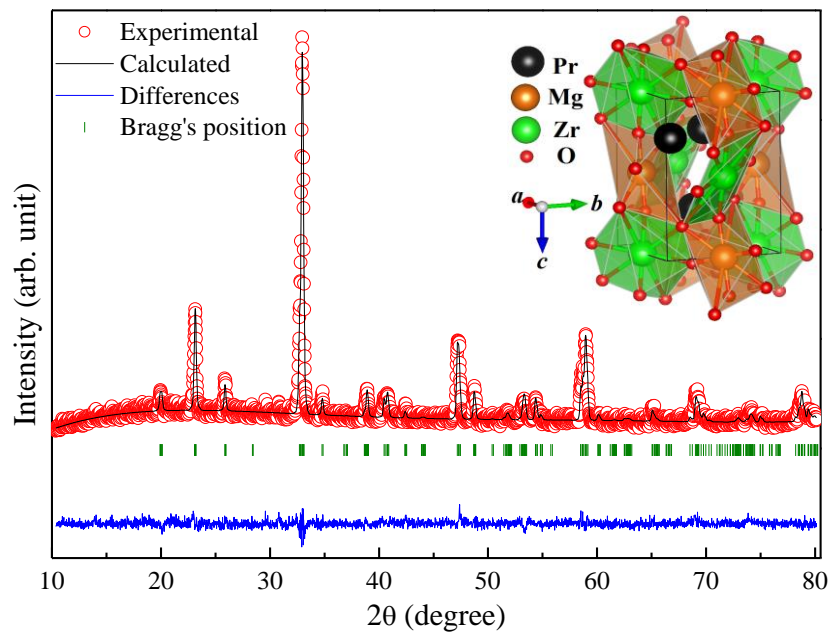

Figure 1: Rietveld analysis results of XRD data for PMZ at room temperature. The inset shows the unit cell of PMZ as obtained from Rietveld analysis.

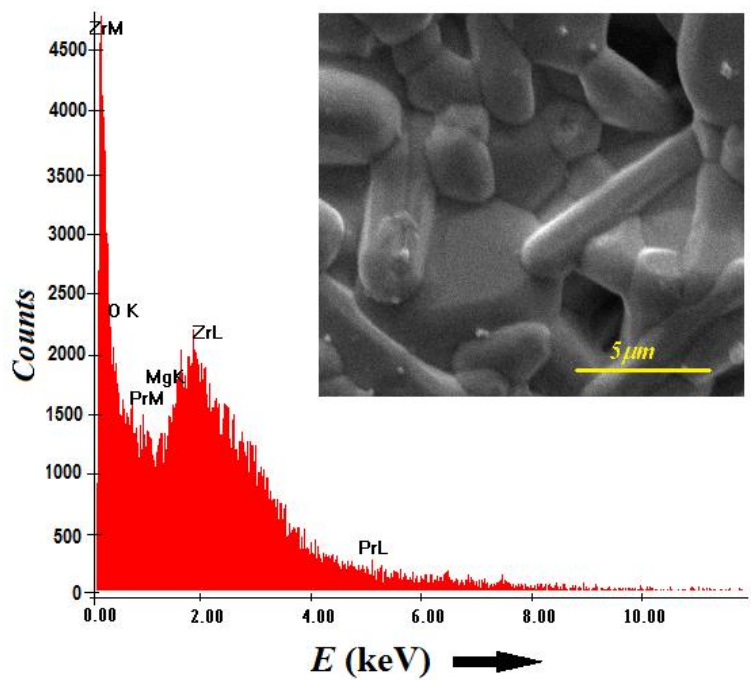

Figure 2: Energy-dispersive X-ray spectra of PMZ. Inset shows the SEM micrograph of PMZ samples.

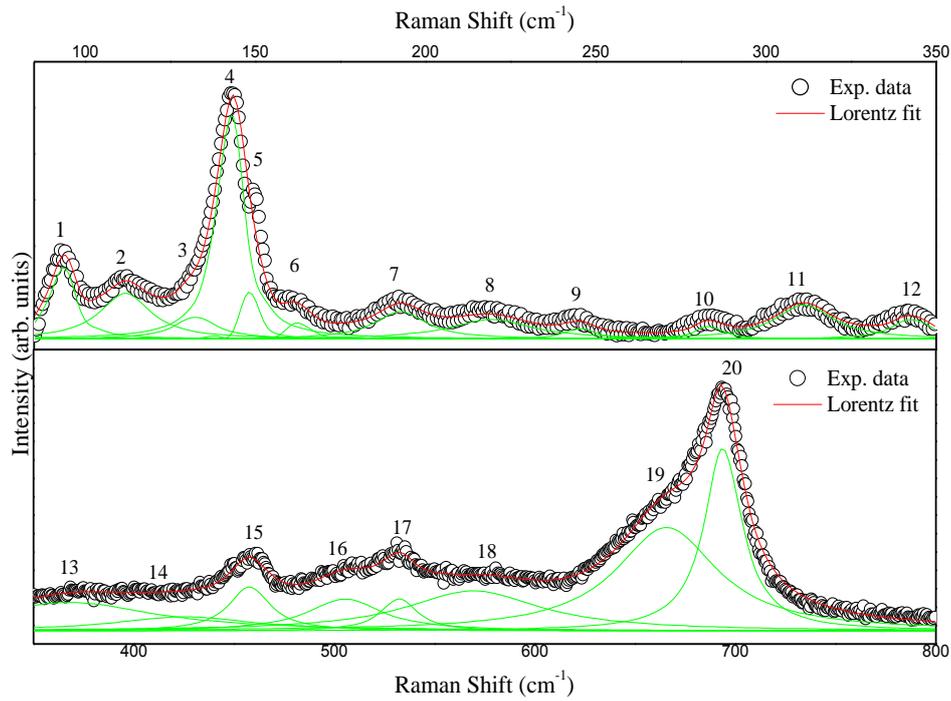

Figure 3: Raman spectrum of PMZ (Experimental data are open circles, while solid lines represent phonon modes adjusted by Lorentzian curves).

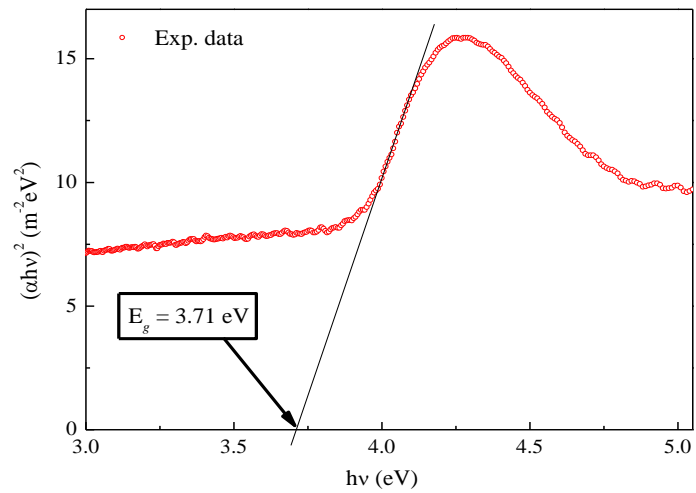

Figure 4: UV-Visible absorption spectrum for direct transition.

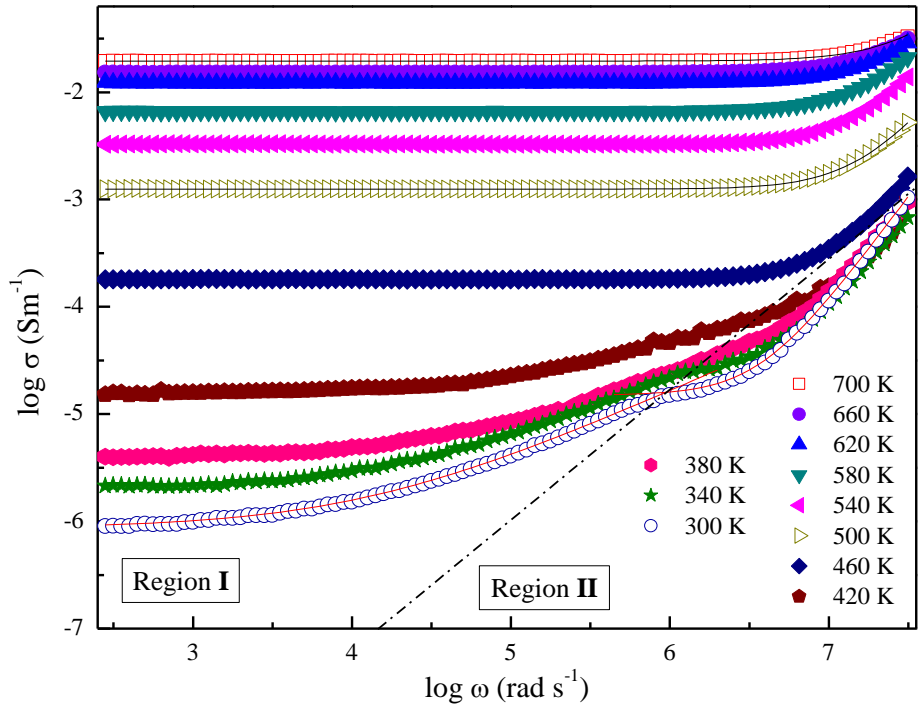

Figure 6: Frequency dependence of the ac conductivity (σ) at various temperatures for PMZ.

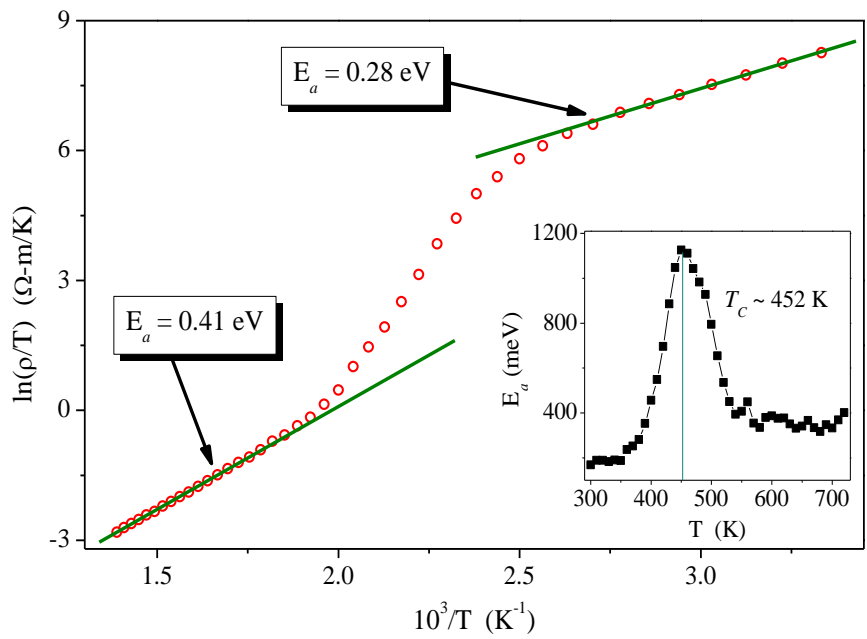

Figure 7: The electrical dc resistivity for PMZ sample. The inset shows the activation energy varies with the temperature.

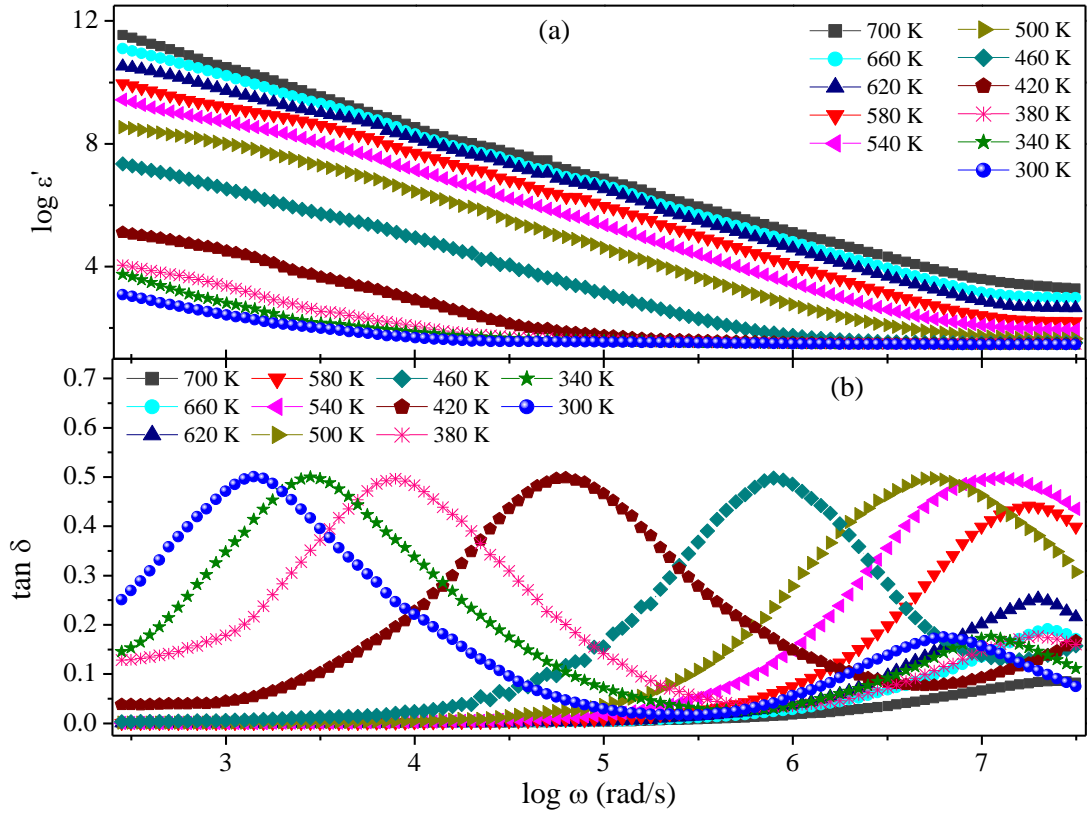

Figure 8: Frequency dependence of log ε′ (a) and tan δ (b) at various temperatures for PMZ.

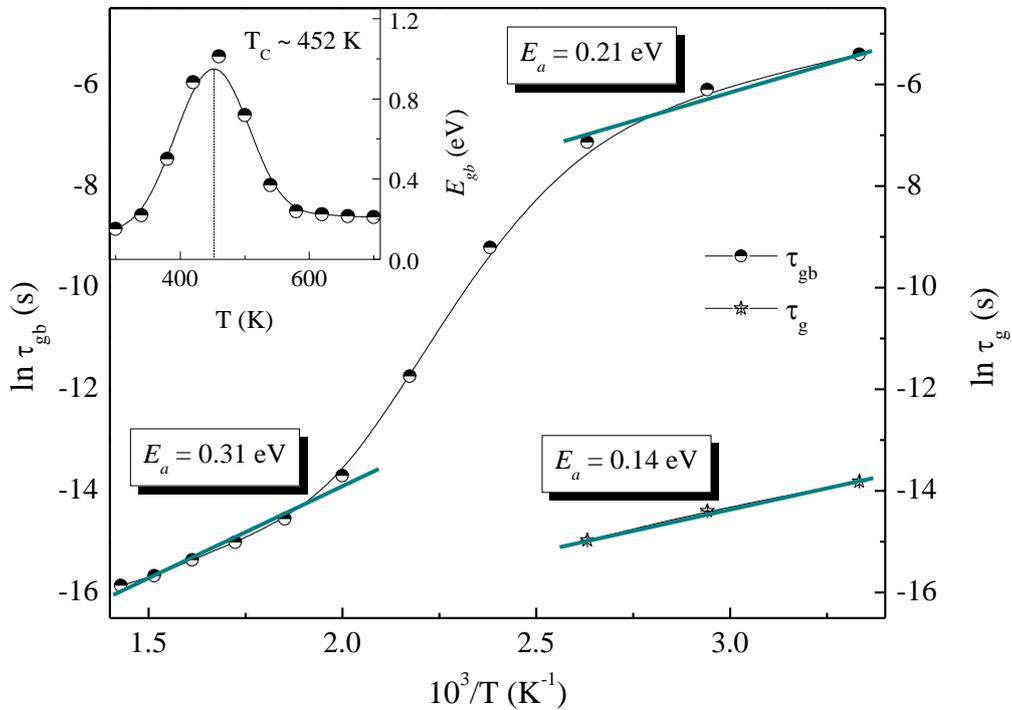

Figure 9: Relaxation time τ of the carriers are plotted against temperature.

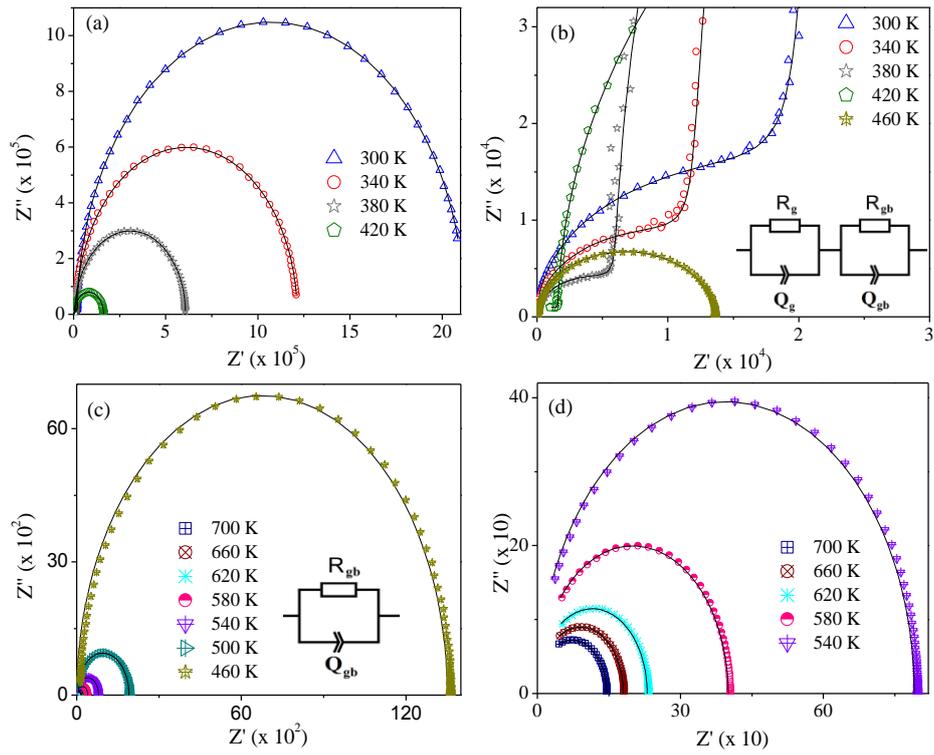

Figure 10: Complex impedance plane plots at various temperatures for PMZ (a, b, c, d). Inset of fig. (b, d) shows the equivalent circuit used to fit the obtained data.

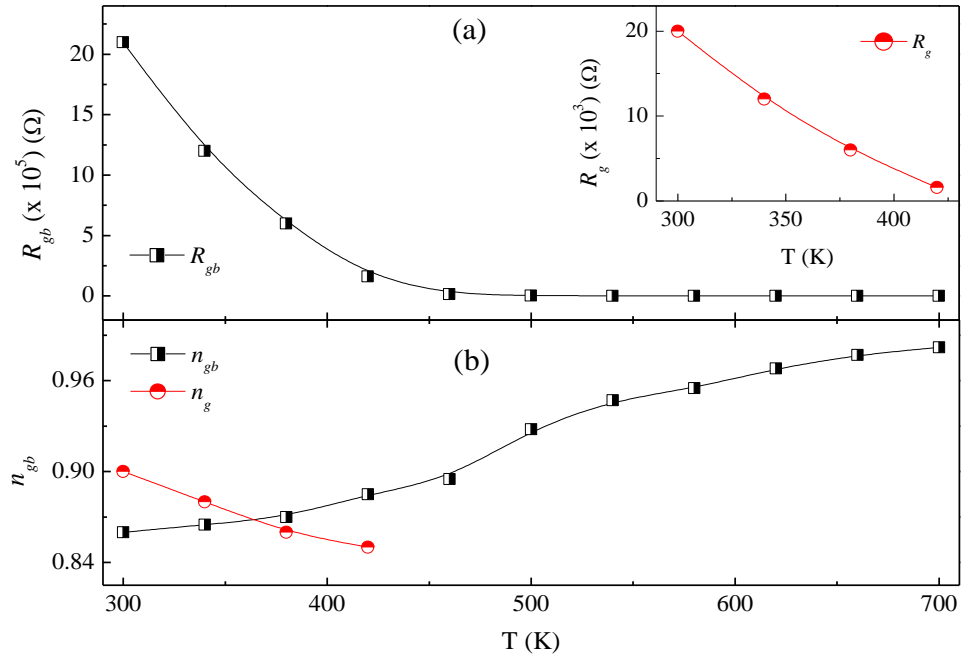

Figure 11: Grain-boundary (a) and grain (inset) resistance varies with the temperature. (c) $n_{gb}$ and $n_g$ plotted with the temperature.

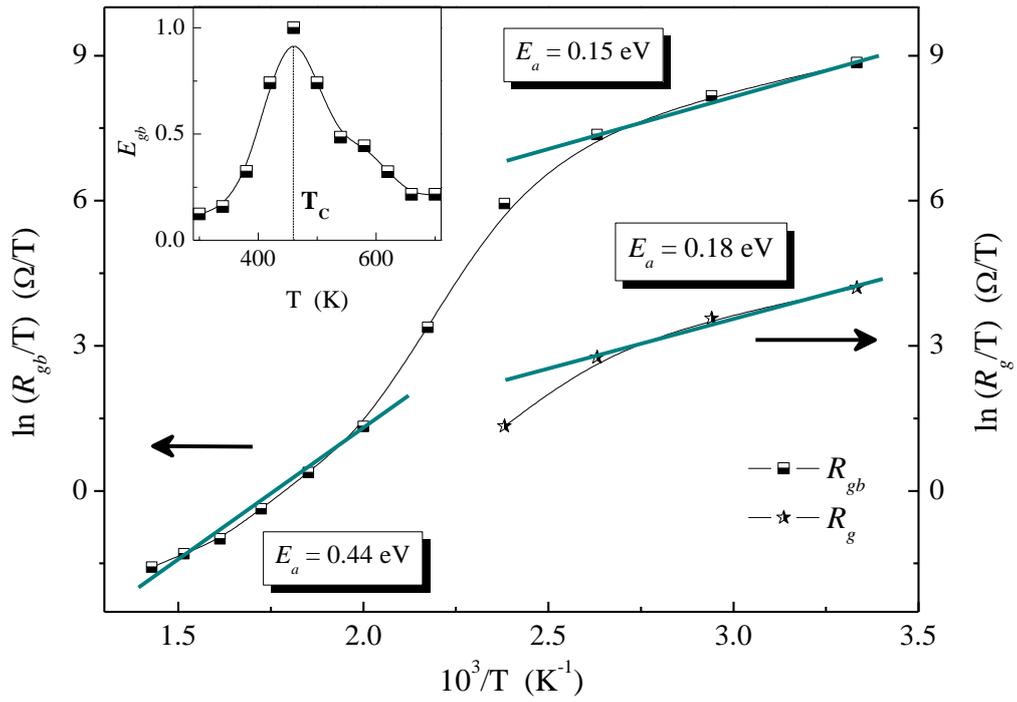

Figure12: (a) Grain-boundary resistance $R_{gb}$ and (b) grain resistance $R_g$ are plotted against the temperature.

# Tables: Exploring the Electrical Transport Properties and Insulator-Metal Transition in Polycrystalline Pr$_2$MgZrO$_6$: Insights from Conductivity and Impedance Spectroscopy

Table 1: Structural parameters extracted from the Rietveld refinement of the XRD data for PMZ at room temperature.

| Space group: P2$_1$/n (Monoclinic) | | | | | | |
|---|---|---|---|---|---|---|
| Cell parameters: $a$ = 5.4166 Å, $b$ = 5.4724 Å, $c$ = 7.6736 Å and $\beta$ = 89.906° | | | | | | |
| Reliability factors: $R_{exp}$ = 7.79, $R_p$ = 6.61, $R_{wp}$ = 8.37 and $\chi^2$ = 1.15 | | | | | | |
| Atoms | Wyckoff site | x | y | z | Bond length (Å) | Bond angle (°) |
| Pr | 4e | 0.49315 | 0.54151 | 0.25600 | Mg–O1 =1.70(8) | Mg–O1–Zr = 148(3) |
| Mg | 2c | 0 | 0.5 | 0 | Mg–O2 = 1.84(5) | Mg–O2–Zr = 155(2) |
| Zr | 2d | 0.5 | 0 | 0 | Mg–O3 = 2.34(5) | Mg–O3–Zr = 105.6(17) |
| O1 | 4e | 0.25457 | 0.33487 | 0.05462 | Zr–O1 = 2.30(7) | |
| O2 | 4e | 0.57822 | 1.00786 | 0.26591 | Zr–O2 = 2.09(5) | |
| O3 | 4e | 0.69429 | 0.28083 | 0.31492 | Zr–O3 = 2.49(5) | |

Table 2: Correlation between gerade modes of the cubic and monoclinic phases of PMZ.

| Cubic $Fm\bar{3}m$ | | Monoclinic $P2_1/n$ |
|---|---|---|
| Free ion symmetry (O$_h$) | Site symmetry (C$_i$) | Unit cell symmetry (C$_{2h}$) |
| ($\nu_1$) $A_{1g}$ | | $A_g$ (3T, 3L, $\nu_1$, 2$\nu_2$, 3$\nu_5$) |
| ($\nu_2$) $E_g$ | $A_g$ | |
| (L) $F_{1g}$ | | $B_g$ (3T, 3L, $\nu_1$, 2$\nu_2$, 3$\nu_5$) |
| (T, $\nu_5$) $F_{2g}$ | | |

Table 3: Observed Raman active phonon modes of PMZ.

| Band no. | Frequency (cm$^{-1}$) | FWHM (cm$^{-1}$) | Symmetry |
|---|---|---|---|
| 1 | 93 | 8 | T |
| 2 | 112 | 18 | T |
| 3 | 132 | 15 | T |
| 4 | 143 | 9 | T |
| 5 | 149 | 5 | T |
| 6 | 163 | 6 | T |
| 7 | 193 | 18 | L |
| 8 | 219 | 29 | L |
| 9 | 245 | 11 | L |
| 10 | 283 | 10 | L |
| 11 | 311 | 21 | L |
| 12 | 343 | 14 | $\nu_5$ |
| 13 | 368 | 105 | $\nu_5$ |
| 14 | 426 | 85 | $\nu_5$ |
| 15 | 458 | 25 | $\nu_5$ |
| 16 | 505 | 45 | $\nu_2$ |
| 17 | 532 | 22 | $\nu_2$ |
| 18 | 569 | 82 | $\nu_2$ |
| 19 | 666 | 61 | $\nu_1$ |
| 20 | 694 | 23 | $\nu_1$ |

Table 4: Fitted parameters for ac conductivity

| Temperature (K) | $\sigma_o$ (S/m) | B | s2 | A | s1 |
|---|---|---|---|---|---|
| 700 | 0.019535 | | | $1.5 \times 10^{-14}$ | 1.6 |
| 660 | 0.015225 | | | $1.0 \times 10^{-14}$ | 1.63 |
| 620 | 0.0122 | | | $6 \times 10^{-15}$ | 1.66 |
| 580 | 0.0066 | | | $2.6 \times 10^{-15}$ | 1.70 |
| 540 | $3.27 \times 10^{-3}$ | | | $0.8 \times 10^{-15}$ | 1.75 |
| 500 | $1.25 \times 10^{-3}$ | | | $1.5 \times 10^{-16}$ | 1.79 |
| 460 | $1.8 \times 10^{-4}$ | | | $1.9 \times 10^{-17}$ | 1.85 |
| 420 | $1.6 \times 10^{-5}$ | $5 \times 10^{-18}$ | 1.9 | $5.7 \times 10^{-9}$ | 0.62 |
| 380 | $4 \times 10^{-6}$ | $9 \times 10^{-18}$ | 1.87 | $3 \times 10^{-9}$ | 0.64 |
| 340 | $2 \times 10^{-6}$ | $8 \times 10^{-18}$ | 1.86 | $2.3 \times 10^{-9}$ | 0.66 |
| 300 | $8.6 \times 10^{-7}$ | $1.25 \times 10^{-18}$ | 1.99 | $1.5 \times 10^{-9}$ | 0.67 |